\documentstyle[eqsecnum,aps]{revtex}

\begin{document}
\draft
\preprint{HEP/123-qed}
\title{Non Metallic Transport in Molecular Solids
versus Dimensionality}
\author{Marco Zoli}
\address{Istituto Nazionale di Fisica della Materia - 
Dipartimento di Matematica e Fisica, \\  Universit\'a di Camerino, 
62032 Camerino, Italy. e-mail: zoli@campus.unicam.it
}

\date{\today}
\maketitle
\begin{abstract}
Path integral techniques and Green functions formalism
are applied to study the
(time) temperature dependent scattering of a polaronic
quasiparticle by a local anharmonic potential in a bath
of diatomic molecules.
The electrical resistivity has been computed in any
molecular lattice dimensionality for different values
of electron-phonon coupling and intermolecular forces. 
A broad resistivity peak with non metallic
behavior at temperatures larger than $\simeq 100K$ is
predicted by the model for sufficiently strong
polaron-local potential coupling strengths. This peculiar
behavior, ascribed to purely structural effects, is favoured in low dimensionality.

\end{abstract}

\narrowtext
\section*{I.Introduction}

A considerable amount of theoretical work has been
devoted to investigate the condition of
polaron formation \cite{eminholst,deraedt,kabanov,kopida,mello,tsiro} 
and the polaronic features
in real materials \cite{kostur,alemott,yarla,devreese}. 
While there is growing evidence
that fundamental properties such as the polaron size,
effective mass and ground state energy are essentially
similar in any dimension \cite{rom,tsiro1}, it is still unclear 
to which extent transport properties in polaronic systems (markedly the
electrical resistivity behavior versus temperature)
depend on the lattice structure and dimensionality. 
Besides being conceptually relevant
this question has become actual in connection with the
discovery of unusual effects in underdoped high $T_c$ superconductors.
Infact the presence of local lattice distortions with
polaron formation has been
envisaged in the high $T_c$ systems and signs of enhanced 
anharmonicity for some in- and out of plane oxygen modes
have been detected in underdoped compounds by several 
groups \cite{miha,palles}.
In this paper we focus on the problem of the interaction
between a polaronic quasiparticle moving through a
molecular lattice and a local structural
instability modelled by a double well potential in its
two state configuration. In particular we derive the
effective interaction strenghts arising from
this peculiar scattering mechanism and we study the effects
on the electrical resistivity both of the intermolecular forces 
and of the lattice dimensionality.

\section*{II. The Model}

Our analysis starts from the following $\tau$ dependent Hamiltonian 
where
$\tau$ is the time which scales as an inverse temperature
according to the Matsubara Green's function formalism:

\begin{eqnarray}
H_{0}(\tau)=& &\, \bar \epsilon(g) \tilde c^{\dag}(\tau) 
\tilde c(\tau) + \sum_{\bf q}{\omega_{\bf q}}a^{\dag}_{\bf q}
(\tau)a_{\bf q}(\tau) + H_{TLS}(\tau) \,
\nonumber \\
& &\Bigl( H_{TLS}(\tau) \Bigr)=\, 
\left(\matrix{0 & \lambda Q(\tau) \cr 
\lambda Q(\tau) & 0 \cr} \right)\,
\nonumber \\  
& &H_{int}(\tau)=\,- 2\lambda Q(\tau) \tilde c^{\dag}(\tau) 
\tilde c(\tau) \,
\nonumber \\
& &Q(\tau)=\, -Q_o + {{2 Q_o}\over {\tau_o}}
(\tau - t_i)
\label{1}.
\end{eqnarray}

$H_0(\tau)$ is the free Hamiltonian made of: a) a polaron
created (distroyed) by $\tilde c^{\dag}(\tau) ( \tilde c(\tau))$ 
in an energy band $\bar \epsilon(g)$ whose width decreases 
exponentially by increasing the strength of the overall
electron-phonon coupling constant $g$, 
$\bar \epsilon(g)=\,Dexp(-g^2)$; b) a lattice of diatomic
molecules whose phonon frequencies
$\omega_{\bf q}$ are derived analytically for a linear chain, 
a square lattice and a simple cubic lattice through a
force constant model; c) a local anharmonic potential
shaped by a Two Level System (TLS) in its 
symmetric ground state configuration.
$Q(\tau)$ is the one dimensional {\it space-time}
hopping path followed by the atom which moves between
two equilibrium positions located at $\pm Q_o$.
$\tau_o$ is the bare hopping time between the two minima 
of the TLS and
$t_i$ is the instant at which the $ith$-hop takes place.
One atomic path is characterized by the number $2n$ of hops, 
by the set of $t_i$ $(0 < i \le 2n)$ and by $\tau_o$.
In the last of eqs.(1), we assume that the class of 
$\tau$-linear paths yields the main contribution to
the full partition function of the interacting system.
The closure condition on the path is given by:
$(2n - 1)\tau_s + 2n\tau_0 =\, \beta$, where  $\beta$ 
is the inverse temperature and $\tau_s$ is the time one 
atom is sitting in a well. 
The interaction is described by $H_{int}(\tau)$ with
$\lambda$ being the coupling strength between TLS and
polaron, 
$\lambda Q(\tau)$ is the renormalized (versus time) tunneling 
energy which allows one to introduce the $\tau$ dependence in 
the interacting Hamiltonian \cite{io3}.

Following a method previously developed 
\cite{hamann} in the study of the Kondo problem, 
we multiply $\lambda Q(\tau)$ by a fictitious 
coupling
constant $s$ $(0 \le s \le 1)$ and, by differentiating with 
respect to $s$,
one derives the  one path
contribution to the partition function of the system

\begin{equation}
ln \Biggl({{Z(n,t_i)} \over {Z_0}} \Biggr)=\, -2 \lambda 
\int_0^1 ds \int_0^{\beta} d\tau Q(\tau) 
 \lim_{\tau^{'} \to \tau^+} G(\tau,\tau^{'})_s  
\label{2}
\end{equation}

where, $Z_0$ is the partition function related to $H_{0}$
and $G(\tau, \tau^{'})_s$ is the full propagator for polarons
satisfying a Dyson's equation:

\begin{equation}
G(\tau, \tau^{'})_s=\,
   G^{0}(\tau, \tau^{'}) + s \int_0^{\beta} dy
   G^{0}(\tau, y) \lambda Q(y)
   G(y, \tau^{'}) 
\label{3}
\end{equation}

The polaronic free propagator $G^{0}$ can be derived exactly
in the model displayed by eqs.(1) \cite{firenze}. 
We get the full partition function of the system  
by integrating 
over the times $t_i$ and summing over all possible even 
number of hops:

\begin{eqnarray}
& &Z_T=\, Z_0\sum_{n=0}^{\infty}
\int_0^{\beta}{{dt_{2n}}\over {\tau_0}} \cdot 
\cdot
\int_0^{t_2-\tau_0}{{dt_{1}}\over {\tau_0}}
exp\Bigl[-\beta E(n,t_i, \tau_0) \Bigr]  \,
\nonumber \\
& &\beta E(n,t_i, \tau_0)=\,
\L -
\bigl(K^{A} + K^{R}\bigr) \sum_{i>j}^{2n}
\biggl({{t_i - t_j}\over {\tau_0}}\biggr)^2 
\label{4}
\end{eqnarray}

with $E(n,t_i, \tau_0)$ being the one path atomic energy. 
$\L$, which is a function of the input parameters,
is not 
essential here while the second addendum in eq.(4) is 
{\it not local in time} as a result of the retarded polaronic
interactions between successive atomic hops in the
double well potential. $K^{A}$ and $K^{R}$ are the one path
coupling strengths containing the physics of the interacting 
system. $K^{A}$ (negative) describes the 
polaron-polaron attraction mediated by
the local instability and $K^{R}$ (positive)
is related to to the repulsive scattering of the polaron
by the TLS. Computation of $E(n,t_i, \tau_0)$ and its
derivative with respect to $\tau_0$ shows that the largest
contribution to the partition function is given by the
atomic path with $\tau_s=\,0$. The atom moving back and
forth in the double well minimizes its energy if it takes
the path with
the highest $\tau_0$ value allowed by the boundary
condition, that is with 
$(\tau_0)_{max}=\, (2nK_BT)^{-1}$.
This result, which is general, provides a 
criterion
to determine the set of dominant paths for the atom at any 
temperature. Then, the effective interaction strengths  
$<K^{A}>$ and  $<K^{R}>$
can be obtained as a function of $T$ by summing 
over $n$ the dominant paths contributions:

\begin{eqnarray}
& &<K^{A}>=\,- {\bigl(\lambda Q_0\bigr)^2 B^2 
exp\bigl({2 \sum_{\bf q}A_{\bf q}}\bigr)
\sum_{\bf q}A_{\bf q} \omega_{\bf q}^2 } 
\tilde f  \,
\nonumber \\
& & <K^{R}>=\,- { \beta \bigl(\lambda Q_0\bigr)^3 B^3 
exp\bigl({3 \sum_{\bf q}A_{\bf q}}\bigr) 
\sum_{\bf q}A_{\bf q} \omega_{\bf q}^2 } 
\tilde f
\label{5}
\end{eqnarray}

with: $B=\,(n_F(\bar \epsilon) - 1) 
exp(- g \sum_{\bf q}ctgh(\beta \omega_{\bf q}/2 ))$ and
$A_{\bf q}=\,2g \sqrt{N_{\bf q}(N_{\bf q} + 1)}$.

$N_{\bf q}$ is the phonon occupation factor and
$n_F(\bar \epsilon(g))$ is the Fermi distribution for
polarons.
$\tilde f =\,\sum_{n=1}^N (\tau_0)_{max}^4$ and
$N$ is the cutoff on the number of hops in a path. 
The particular form of
$(\tau_0)_{max}$ suggests that many hops paths
are the relevant excitations 
at low temperatures whereas paths with a low number
of hops provide the largest contribution to the 
partition 
function at high temperatures. Since the effective
couplings which determine the resistivity depend 
on $(\tau_0)_{max}^4$, hence
on $N^{-3}$ (through $\tilde f$), a
relatively small cutoff ($N \simeq 4$) ensures numerical 
convergence of eqs.(5) in the whole temperature range.
On the other hand the non retarded term $\L$ in eq.(4) 
has a slower
$1/N$ behavior,
therefore a larger cutoff should be taken
at low temperatures where computation of
equilibrium properties such as specific heat is strongly 
influenced by many hops atomic paths between the minima
of the double well excitations.

The lattice Hamiltonian is made of diatomic sites
whose intramolecular vibrations can favor trapping
of the charge carriers  \cite{holstein}. 
The {\it intra}molecular 
frequency $\omega_0$ largely determines the size of
the lattice distortion associated with  polaron formation
\cite{alenew} while the dispersive features of the phonon
spectrum controlled by the first ($\omega_1$) and second
($\omega_2$) neighbors {\it inter}molecular couplings
are essential to compute the polaron properties
both in the ground state and at finite temperatures
\cite{io2}.
The range of the intermolecular forces is extended to the
second neighbors shell since these couplings 
remove the phonon modes
degeneracy (with respect to dimensionality) at the
corners of the Brillouin zone thus permitting to estimate 
with accuracy the relevant contributions of high symmetry 
points to the momentum space summations \cite{io1}.
Then the characteristic frequency $\bar \omega$, 
which we choose as the zone 
center frequency, is: 
$\bar \omega^2=\, \omega_0^2 + z\omega_1^2 
 + z_{nnn}\omega_2^2$, $z$ is the coordination number and 
$z_{nnn}$ is the next nearest neighbors number.
Hereafter we take frequencies values which are appropriate
to systems with rather sizeable phonon spectra.

We turn now to compute the electrical resistivity 
due to the polaronic charge carrier scattering by
the impurity potential with internal degree of freedom
provided by the TLS.
Assuming s wave scattering, one gets \cite{mahan,yuand}

\begin{eqnarray}
& &\rho=\, \rho_0 sin^2 \eta \,
\nonumber \\
& &\rho_0=\, {{3 n_s} \over {\pi e^2 v_F^2 
\bigl( N_0/V \bigr)^2 \hbar}}\,
\label{6}
\end{eqnarray}

where, $n_s$ is the density of TLS which act as scatterers,
$v_F$ is the Fermi velocity, $V$ is the cell volume, 
$N_0$ is the electron density of states,
$e$ is the 
charge and $\hbar$ is the Planck constant. The 
phase shift $\eta$ to the electronic wave function at the Fermi 
surface
is related to the effective interaction strengths 
$<K^{A}>$ and  $<K^{R}>$.

The input parameters of the model are six that is,
the three molecular force constants, $g$, $D$ and the bare
energy $\lambda Q_0$.
$Q_0$ can be chosen as $\simeq 0.05 \AA$ consistently with
reported values in the literature on TLS's which are
known to exist in glassy systems 
\cite{ander,phil,fessa,zawa}, 
amorphous metals \cite{coch},
A15 compounds and likely in some cuprate 
superconductors \cite{mustre,saiko,menu}. In these systems the 
origin
of the TLS's is not magnetic. The bare electronic band
$D$ is fixed at 0.1eV.

Let's start by studying (Fig.1) the resistivity behavior
in a 3D lattice as a function of the overall electron-phonon 
coupling $g$.
$\lambda$ is fixed at $700meV \AA^{-1}$ which means a
bare tunneling energy of $\simeq 35meV$ and the molecular
force constants are $\omega_0=\,60meV$, $\omega_1=\,50meV$ 
and $\omega_2=\,20meV$, respectively.
At low temperatures
$\rho(T)$ approaches the unitary limit for any of the four
$g$ values thus displaying the peculiar effect of the
anharmonic potential with internal degree of freedom.
At $T \simeq 150K$, $\rho$ develops a $g$-dependent
broad peak which softens and finally disappears
by increasing $g$ in the strong coupling regime. We note
that at $g > 1$ the polaron energy becomes smaller than 
the atomic tunneling energy hence polaron scattering by
the TLS is essentially diagonal. On the contrary, in the
intermediate coupling regime ($g$ in the range 0.5 - 1) 
the incoming
polaron can release a sufficiently high amount of energy,
off diagonal scattering by the impurity potential prevails
and a broad resonance peak emerges. At weaker $g$ the
polaronic picture would lose any validity.

Next we turn to consider the effect of the lattice
dynamics on the $\rho$ versus $T$ behavior. In Figs.2 and
3, the normalized resistivities are reported on for
the 1D and 2D molecular lattices, respectively. 
Let's set $g=\,1$ which ensures
both polaron formation and presence of the resonance peak
also in 1D and 2D systems while the tunneling energy 
(as in Fig.1) is in the range of the values estimated by
EXAFS investigations in high $T_c$ systems with double
site distributions for the 
apical oxygen atoms \cite{mustre,haskel}. 
By taking $\omega_0=\,60meV$ and $\omega_1=\,50meV$ we choose
for the first neighbors intermolecular coupling model a 
characteristic phonon energy $\bar \omega \simeq 0.1eV$. 
This choice
allows us: i) to treat correctly the ground state polaron
properties versus dimensionality \cite{io1}, ii) to reproduce
the sizable phonon energy of c-axis polarized mode due to
apex oxygen vibrations coupled to the holes in the Cu-O planes of $YBa_2Cu_3O_{7-\delta}$. In underdoped compounds (whose
c-axis resistivity displays the unusual non metallic behavior)
this mode appears enhanced in energy \cite{timusk}. 
Although our simple cubic lattice does not account
for the details of the structural effects of $YBCO$ we  
are therefore in the appropriate range of parameters to 
capture the main features of the
lattice polarons scattered by local instabilities in those compounds.
The role of the second
neighbors couplings is emphasized in Figs.2 and 3
by varying $\omega_2$ 
which however should not exceed $\omega_1$.
Increasing $\omega_2$ the polaron spreads in real space
and becomes lighter. Accordingly the height of the peak
decreases. We also note that the strength of the
intermolecular forces influences the position of the peak
versus temperature which results from a balance between
competing attractive and repulsive interactions (eqs.(5)).
At fixed parameters, the height of the 1D peak is roughly
twice as much that of the 2D one. Comparing the dotted
2D curve ($\omega_2=\,20meV$) with the corresponding
case ($g=\,1$ plot) in Fig.1, one sees that the 3D peak
is further reduced and its absolute value lies below
the residual resistivity value.

The dynamics of the TLS and its coupling to the
charge carrier strongly affect the transport in any
dimensionality as Figs.4 make evident. When $\lambda$
is small (i.e. below $300meV \AA^{-1}$ in 1D) 
polaron and TLS are weakly coupled, off diagonal scattering 
is unlikely to occur and the conductivity is metallic-like.
Increasing $\lambda$, the atomic tunneling energy
becomes of order of the polaron energy and the conditions
for resonant scattering are established. Note that by
varying $\lambda$ the resistivity peak does not shift
versus temperature while   
the $\lambda$ threshold value for the appearance of
metallic conductivity is larger in higher dimensionality.
This means that 3D systems can sustain a larger degree
of anharmonicity (than low dimensional systems) and still
exhibit metallic transport properties while trapping of
the charge carrier by the anharmonic impurity potential is 
favoured in 1D. It is this trapping which causes the
broad resistivity peak and non metallic transport at
$T$ larger than $\simeq 100K$ in 1D.

We investigate further the effect of the molecular forces
(Figs.5) by {\it increasing} the {\it intra}molecular energy
and {\it decreasing} the {\it inter}molecular energies
with respect to Figs.4.
The 1D peak (compare Fig.5(a) and Fig.4(a))
is reduced for any $\lambda$ value whereas the 2D and
3D peaks are here larger than in Figs. 4(b) and 4(c)
respectively. The reason is the following: in 1D, the
intramolecular coupling is determinant because of
the low coordination number; as a result, by enhancing
$\omega_0$ the carriers become lighter and the absolute
resistivity is reduced. On the contrary in 2D and mostly
in 3D, long range effects are more effective hence, the
stronger $\omega_0$ (with respect to Figs.4) is more than
balanced by the weaker $\omega_1$ and $\omega_2$; as a
result, the characteristic 2D and 3D phonon frequencies
are smaller, polarons become less mobile and the
resistivity peaks are accordingly higher. 

\section*{III. Conclusion}

The path integral formalism has been applied to study
the time retarded interacting problem of a polaron
scattered by a local anharmonic potential in a lattice
of diatomic molecules. I have derived
the full partition function of the system and obtained
analytically the effective coupling strengths as a function
of temperature. The electrical resistivity has been computed
in any lattice dimensionality for a large choice of input parameters.
Strength of the overall electron-phonon coupling ($g$),
strength of the polaron-local potential coupling ($\lambda$)
and strength of the molecular forces interfere, giving rise to
a rich variety of resistivity behaviors versus temperature.
As a main feature, when the conditions for resonant scattering 
between polaron and
local double well potential are fulfilled, a broad resistivity peak
shows up in the 100-150K range. 
However, the shape of the peak essentially
depends on the dynamics of the local potential (tuned by $\lambda$)
and, for sufficiently low atomic tunneling energies, metallic
conductivity conditions are recovered.
Generally, in 3D the resistivity maximum is less pronounced than
in low dimensionality and, for sufficiently strong intermolecular
couplings, the height of the peak lies below the residual
resistivity value, being therefore hardly visible in experiments.
On the other hand, in 1D and 2D, the maximum can easily become 
larger than the residual
resistivity by a factor $\simeq 3-4$ for realistic choices of
tunneling energies and lattice force constants. Then, non metallic
resistivities can be expected in low dimensional polaronic systems
with local lattice instabilities at least for temperatures 
above $\simeq 100K$. 
As a final remark one should add that, 
while the {\it intra}- and {\it inter}molecular energies have been
varied throughout the paper as apparently independent parameters,
their values for real systems can be obtained (for instance)
by a least squares fitting procedure to experimentally known 
physical quantities \cite{io5}.

\begin{figure}
\vspace*{6truecm}
\caption{Electrical Resistivity normalized to the residual
($T=\,0$) resistivity for four values of electron-phonon coupling
$g$. The bare TLS energy $\lambda Q_0$ is 35meV. 
The force constants which control the phonon
spectrum in the 3D lattice are: $\omega_0=\,60meV, \omega_1=\,50meV, \omega_2=\,20meV$}
\end{figure}

\begin{figure}
\vspace*{6truecm}
\caption{1D Electrical Resistivity  as a function of the 
second neighbors 
intermolecular force constant.  $g=\,1$.
$\lambda Q_0$, $\omega_0$ and $\omega_1$ are as in Fig.1}
\end{figure}

\begin{figure}
\vspace*{6truecm}
\caption{2D Electrical Resistivity as a function of the 
second neighbors 
intermolecular force constant.  $g=\,1$.
$\lambda Q_0$, $\omega_0$ and $\omega_1$ are as in Fig.1}
\end{figure}

\begin{figure}
\vspace*{6truecm}
\caption{Electrical Resistivities for six values of the 
polaron-TLS coupling
$\lambda$ in units $meV \AA^{-1}$. $g=\,1$.  $\omega_0=\,60meV, \omega_1=\,50meV, \omega_2=\,30meV$.
(a) one dimension; (b) two dimensions; (c) three dimensions}. 
\end{figure}

\begin{figure}
\vspace*{6truecm}
\caption{ Electrical Resistivities for six values of the 
polaron-TLS coupling
$\lambda$ in units $meV \AA^{-1}$. $g=\,1$.  $\omega_0=\,80meV, \omega_1=\,40meV, \omega_2=\,20meV$.
(a) one dimension; (b) two dimensions; (c) three dimensions}. 
\end{figure}

\section*{Acknowledgements}

I'm thankful to V.R.Sobol and O.N.Mazurenko for helpful
collaboration. A NATO-CNR grant is acknowledged.

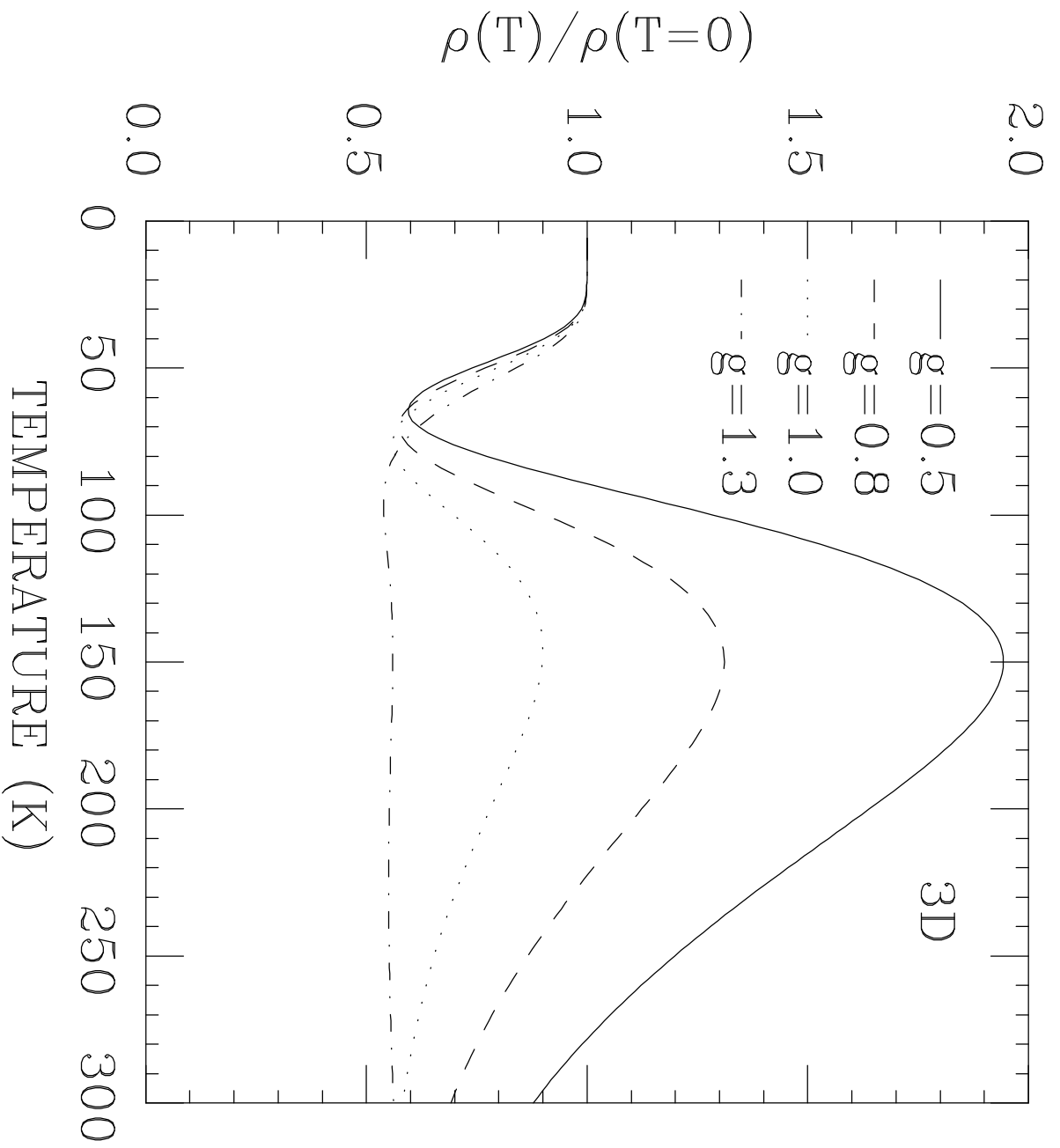
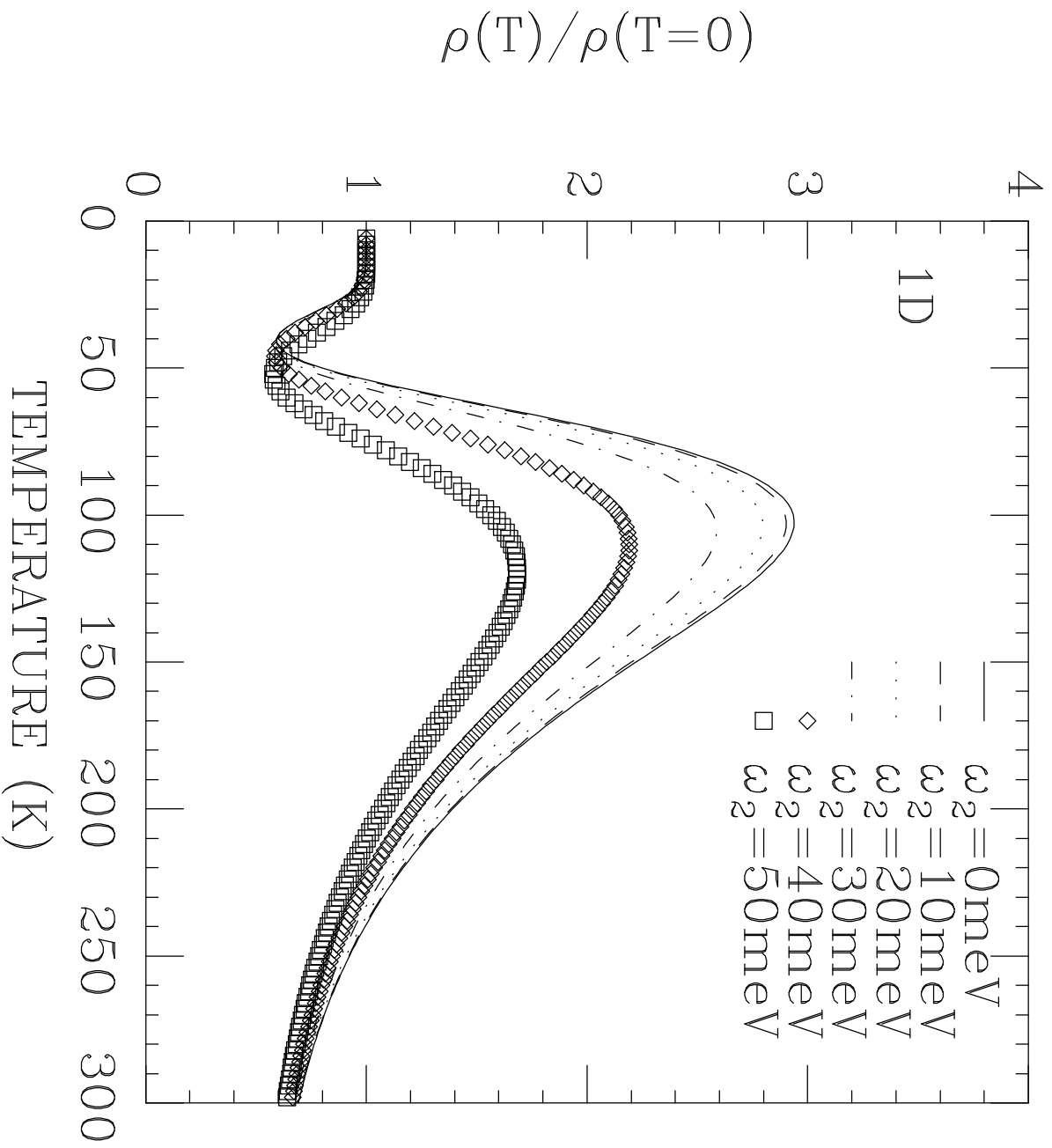
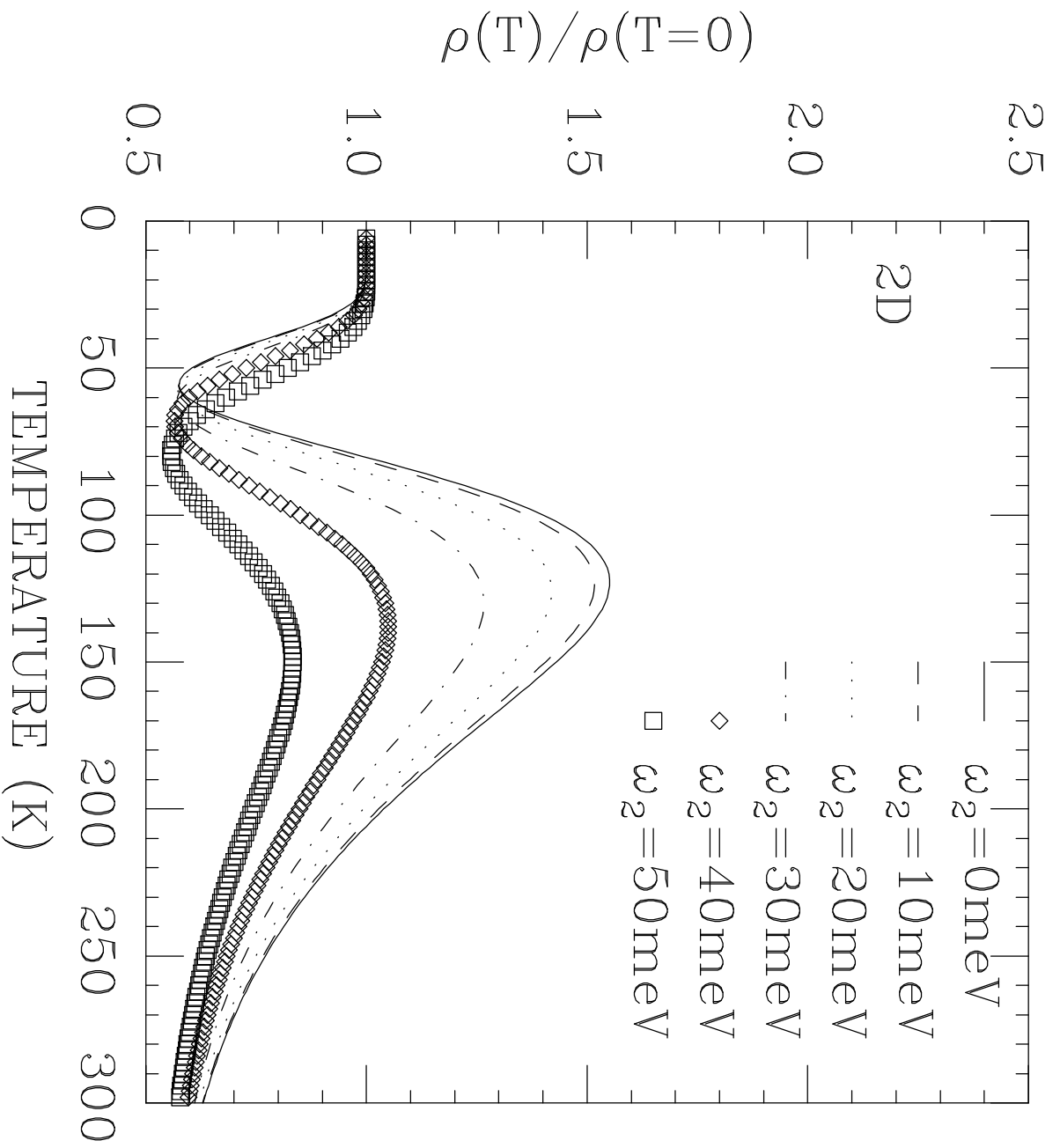
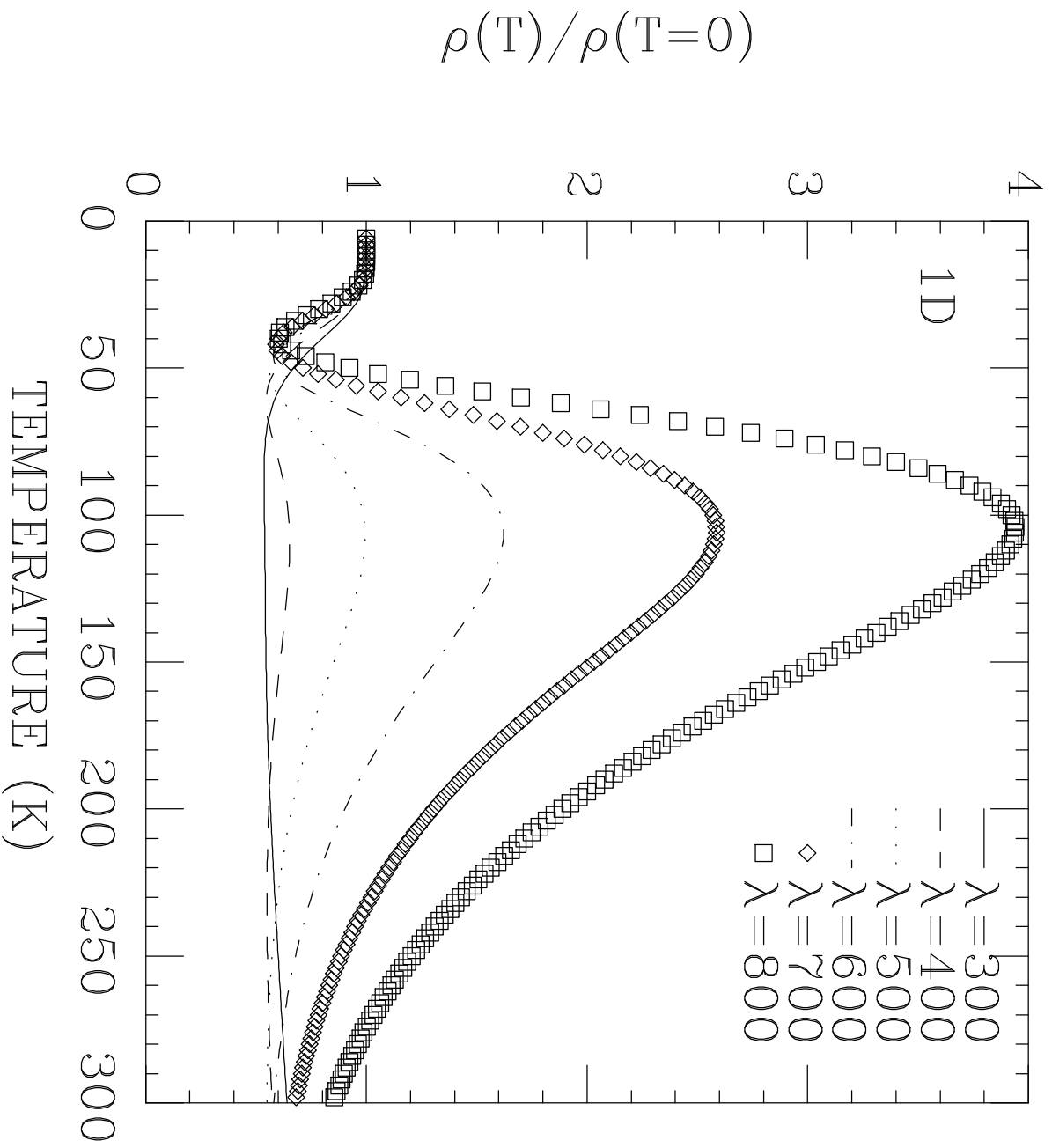
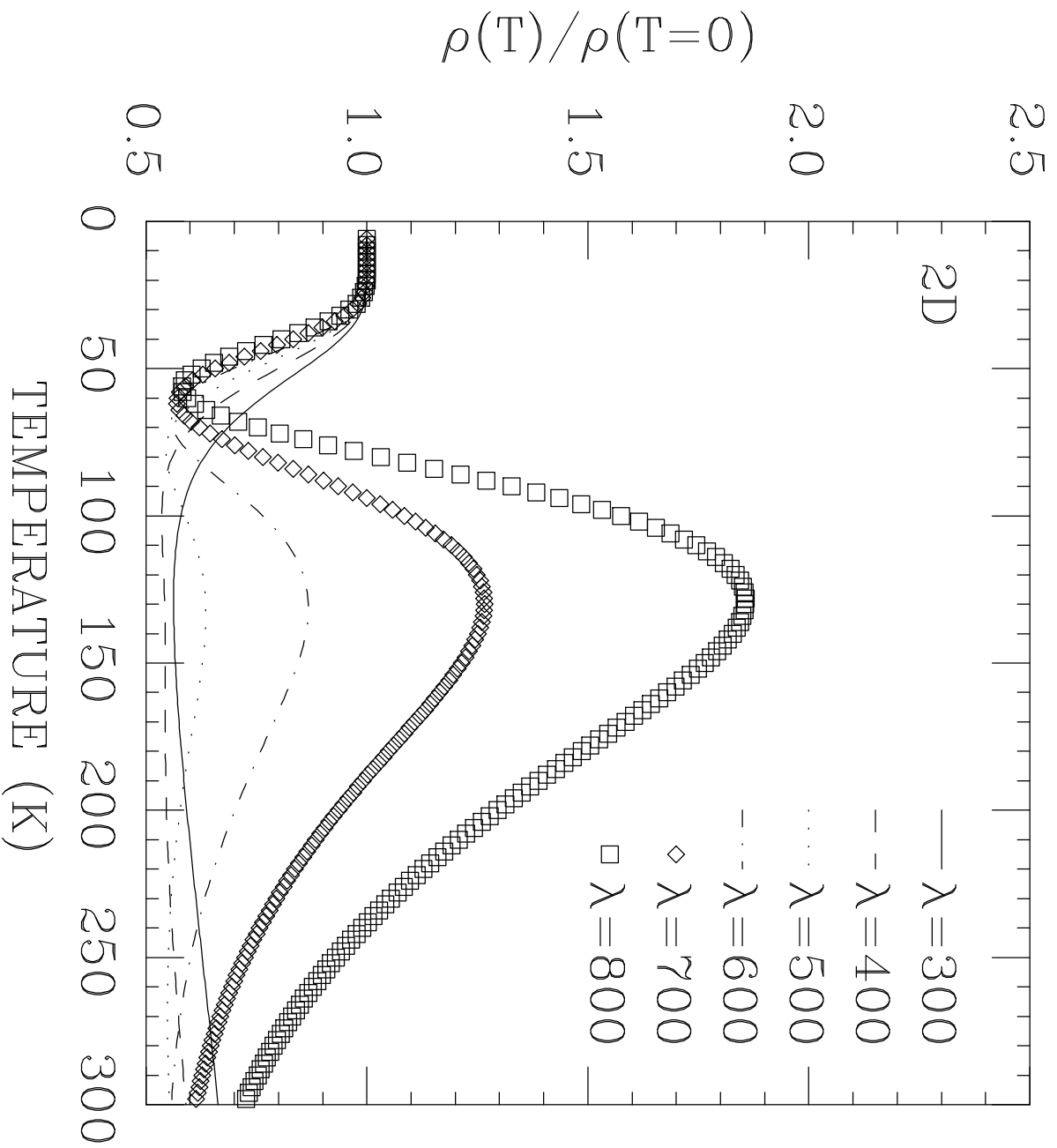
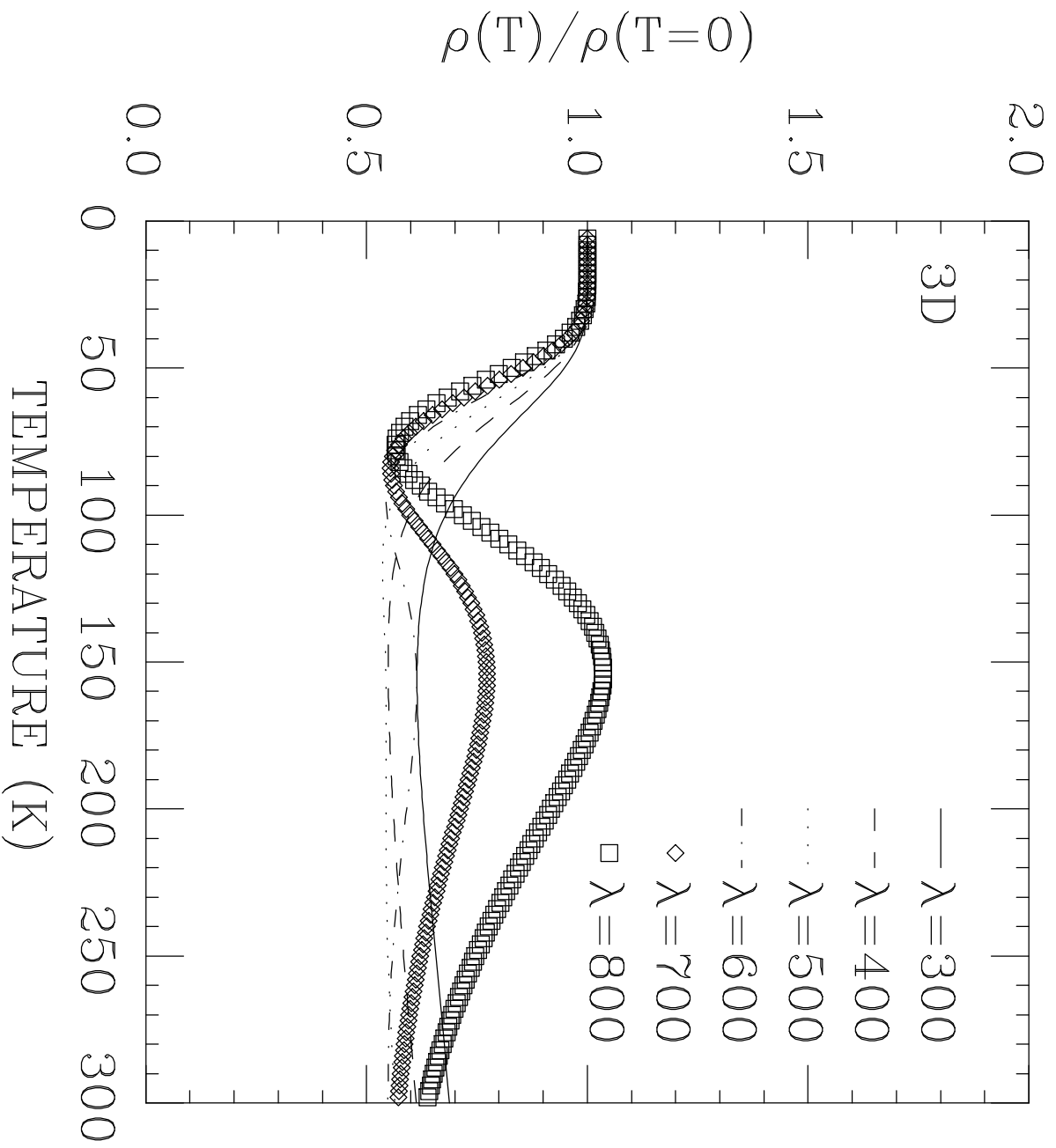
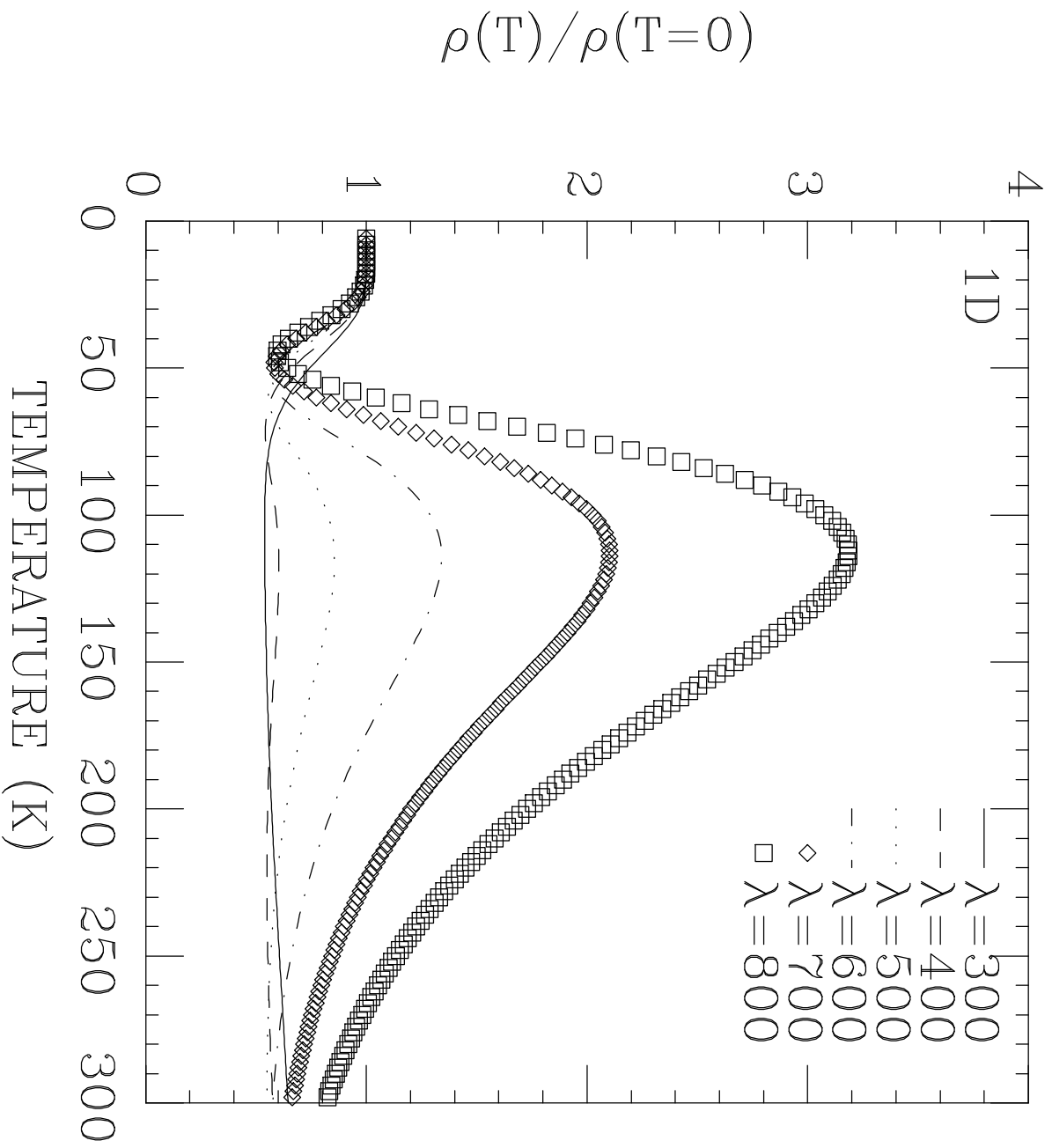
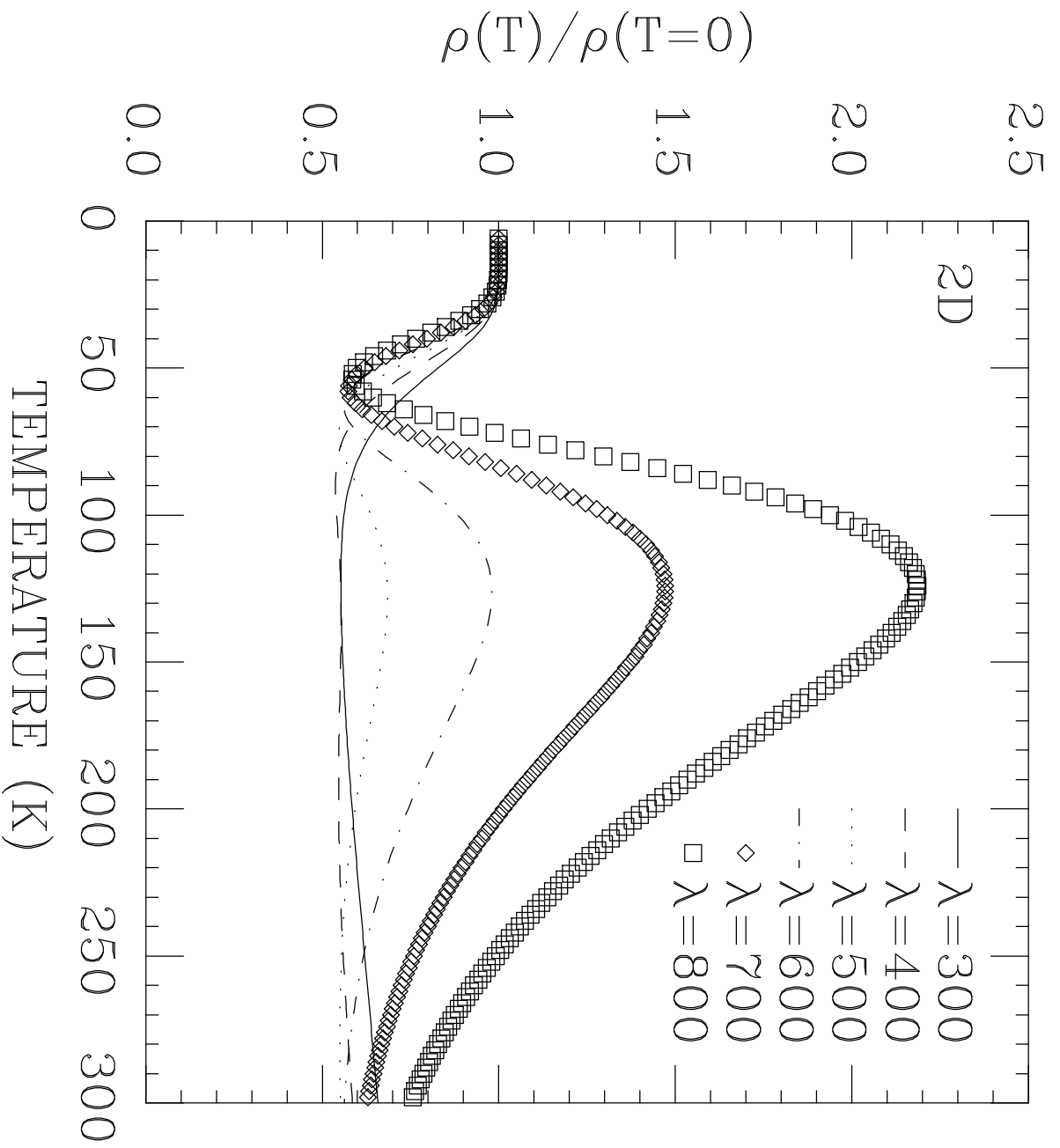
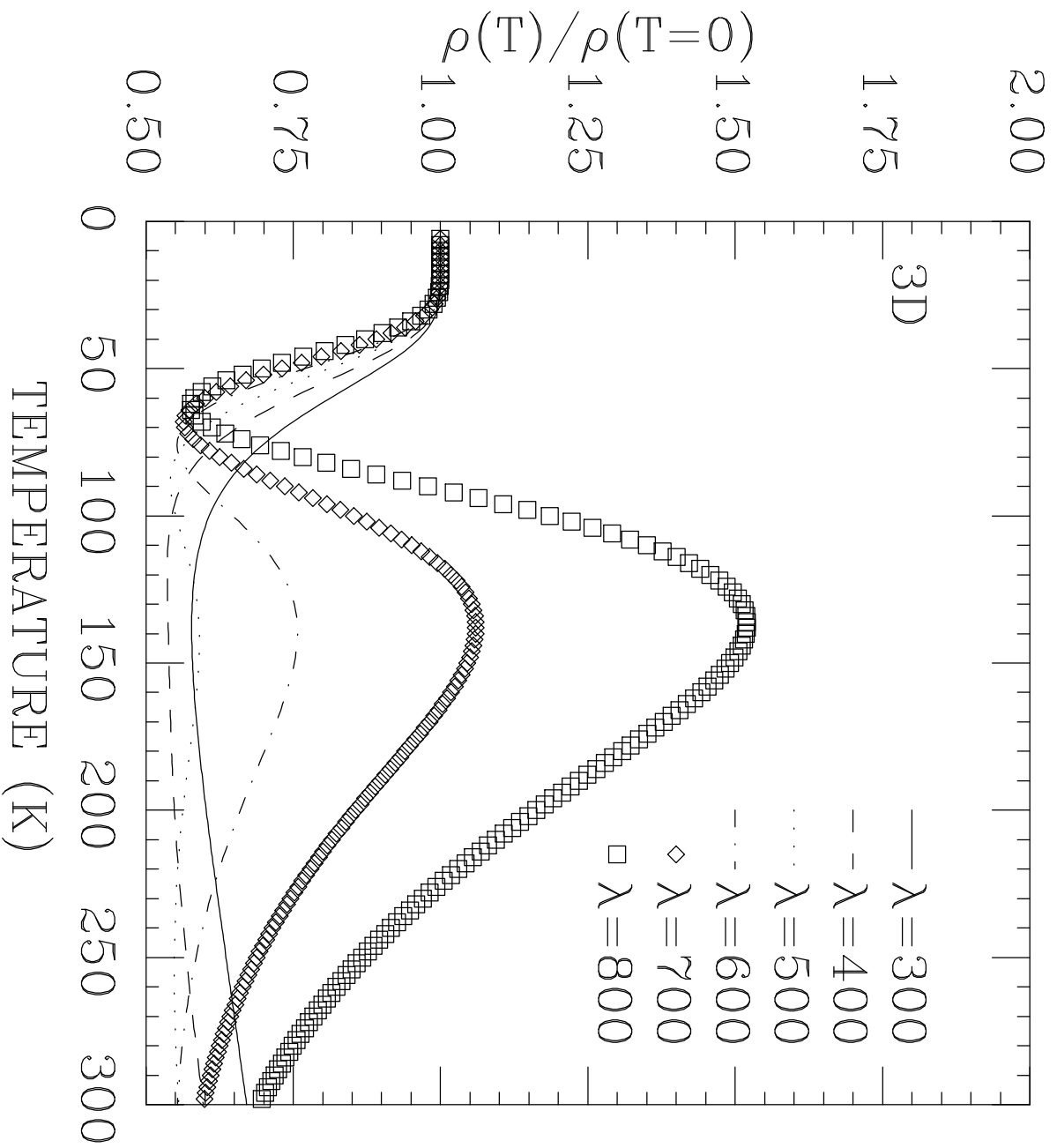
\end{document}